\documentclass{ ametsocV6.1}
\setcitestyle{numbers} 
\renewcommand{\linenumbers}{}
\renewcommand{\internallinenumbers}{}
\renewcommand{\modulolinenumbers}[1]{}

\usepackage{amsmath}
\usepackage{amsfonts} 
\usepackage{amssymb}

\newcommand{\mbf}[1]{\mbox{\boldmath $#1$}}

\def\0{\mbox{\mbf 0}}

\def\bx{\mbox{\mbf x}}

\def\BibTeX{{\rm B\kern-.05em{\sc i\kern-.025em b}\kern-.08em
    T\kern-.1667em\lower.7ex\hbox{E}\kern-.125emX}}

\usepackage{graphicx}

\usepackage{dcolumn}

\usepackage{bm}

\usepackage{amsmath}

\title{Is the Quantum-Entangled Universe a Small World?}

\author{Gregory S. Duane\footnote{Corresponding Author: gregory.duane@colorado.edu 
303-492-7263  fax: 303-492-3524}}
\affiliation{%
Dept. of Atmospheric and Oceanic Sciences, UCB 311, University of Colorado, Boulder, CO 80309 }

\abstract
  {Partial entanglement may provide enough connections in the universe to satisfy the definition of a small world. To investigate this possibility, we define a network of particles on a given space-like hyper-surface with a long-range link between any two particles that are connected by a sequence of exchanged particles  involving a limited number of interactions. Considering the mean free paths of particles in different regions of space and the resulting probability distributions of entanglement connections vs. distance, we find evidence of small-world or random network structure on all but the smallest scales  - corresponding to stars and planets - on which other types of connections would need to be added to complete the small world picture.}

\AtBeginDocument{%
\nolinenumbers
}
\begin{document}
\maketitle

{\bf That a few long-range connections can drastically reduce the average length of the shortest path between any two nodes in a network whose nodes are otherwise only locally connected is a surprising phenomenon underlying the well-known six degrees of separation characterizing connections in human society. We explore the question of whether the physical universe could similarly be described as a small world. The long-range connections are provided by pairs of quantum-mechanically entangled particles, which in empty space can migrate to opposite sides of the universe, yet remain intimately connected in the sense that a measurement or interaction of one immediately provides information about the state of the other. In a preliminary approximate treatment, we show that on scales larger than those of stars and planets, the probability of partial entanglement falls with distance slowly enough that small-world structure is maintained – the spatially largest known example of this phenomenon to date. Consequences remain to be determined. One possible consequence is that synchronized phenomena on more familiar scales, if connected through entanglement to processes elsewhere, could give rise to synchronized phenomena on a cosmic scale.}

\section{Introduction}

The existence of entangled particle pairs, separated by arbitrarily large distances, introduces connections in the universe not present in the classical world view.  Long-range connections in a network can render the network a ``small world". in which the average degree of separation of any two nodes is markedly reduced. It is therefore natural to ask whether entanglement has such an effect on the structure of the world, and what might be the consequences. In this short paper we address primarily the first question, exploring conditions under which a naive distribution of entangled particle pairs would give rise to small world structure. We find evidence that the known non-uniform  distribution of matter in the universe gives rise to small-world or random network structure at all but the  smallest scales.

By way of motivation, we note that small-world structure facilitates synchronisation of dynamical elements at the nodes of the network. One does not need to invoke faster-than-light message passing to imagine that maximally entangled particle pairs could support long-distance synchronisation. The impact of entanglement would thus be multiplied, with consequences well beyond what can be discussed here.

Previous authors have investigated the propagation of entanglement itself in networks that are {\it constructed} so as to have small-world structure \cite{refAbedi, refBrito}.  Such networks could be engineered for practical applications such as secure communications  and network robustness. It has also been suggested that entanglement could give rise to space-time and its geometry \cite{refMaldacena}, thus laying a new basis for physics.  But to our knowledge, the question of whether there is small-world structure in the universe with geometry and distribution of matter as known or conjectured has not previously been addressed.

We begin in the next section by sketching the requirements for small-worldedness, as applied to a network with connections defined so as to include those given by entanglement. In Section 3, we compare these requirements to known cosmography
to infer small-world structure in large regions of the universe.
Finally, in the concluding section, we summarize the extensions of this work and further definitions that will be needed to ascribe small-world structure to the universe as a whole.

\section{Network Structure}
\label{sec2}

Putting aside for now the implications of connection via entanglement, we seek to define the network structure that corresponds to the intuitive notion that particle pairs entangled across large distances render the universe a small world. Toward this end, we first recall the general definition of a small-world network as one in between a random network and a locally connected network.
In a random network, there is no locality at all in the connection pattern; all pairs of nodes have an equal chance of being connected, making the average path length - the average of the minimum number of links required to connect any two nodes - a small number that does not grow in proportion to the size of the network. In a locally connected network the average path length does grow in proportion to network size.  In a small-world network, one preserves some notion of local connection via a requirement of ``clustering" \cite{refWattsStrogatz},  but introduces a small number of long-range connections that give short average path lengths which grow only logarithmically with network size.

One way to realise small-world structure, which we will rely on here, is to posit an initial metric with respect to which most  connections are local. That is easily achieved if we imagine that the nodes reside on a rectangular lattice of some given low dimensionality $k$, with the distance between any two nodes on the lattice taken as the ``Manhattan" distance in Kleinberg's original construction  \cite{refKleinberg}.
We then imagine that each node is connected to all its nearest neighbours, except for a small fraction $p$ of connections assigned to other nodes,  randomly chosen in the network. The resulting network can be shown to be small-world, even for small $p$, if the network is large enough that there is a significant probability of connection to very distant nodes \cite{refBarthelemy}.  A further modification is to make the probability of connection to another node depend on the distance to that other node, for some general distance metric.  Small-worldedness then depends on the form of the probability distribution. For a network with connections to a given node that initially include only nearest neighbours  on the $k$-dimensional lattice, with random added wiring to distant nodes based  on a power-law dependence of the probability $p$ of such wiring on distance $d$, i.e.  $p_{AB} \propto d(A,B)^{-r}$, Nguyen and Martel \cite{Nguyen, NguyenExt} showed that the network is small-world if and only if 
\begin{equation}
\label{Nguyenbound}
k < r < 2k 
\end{equation}
Larger values of $r$ result in too few long-range connections for a small world, while smaller values allow single connections that are too long to maintain clustering.

Here, we consider a network of elementary particles, say those of the Standard Model, on a space-like hypersurface with an initial metric given by physical distance. 
The initial connection scheme posits a link between any pair of nodes separated by a distance less than some value $D$.  At each node, long-range connections are added that connect to any particle with which the particle at the first node has some minimum degree of entanglement (defined in Appendix \ref{AppendixA}).  

\begin{figure}
a) \resizebox{0.6\textwidth}{!}{\includegraphics[angle=0, trim = 0.5in  3in 4in 2in, clip]{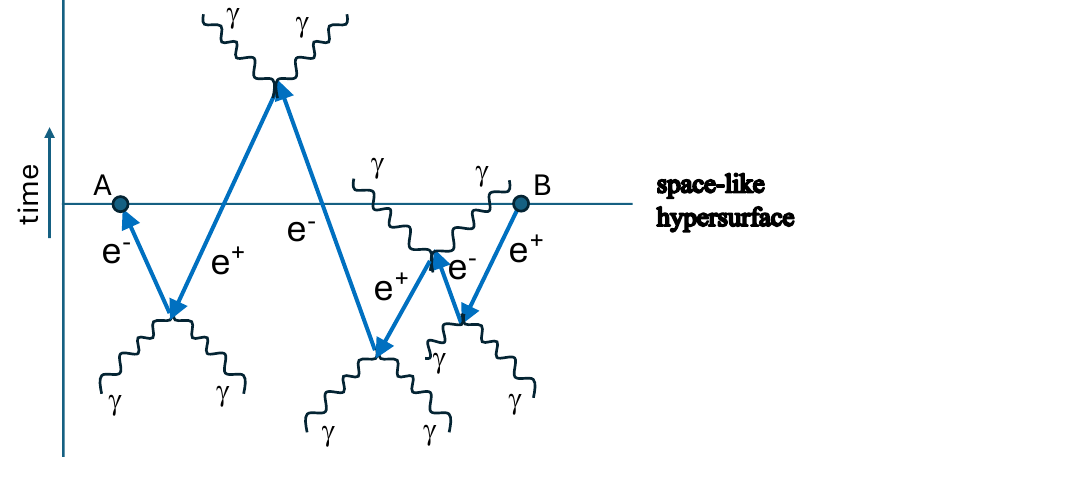}} \\
b) \resizebox{0.6\textwidth}{!}{\includegraphics[angle=0, trim = 0.5in  2in 4in 1in, clip]{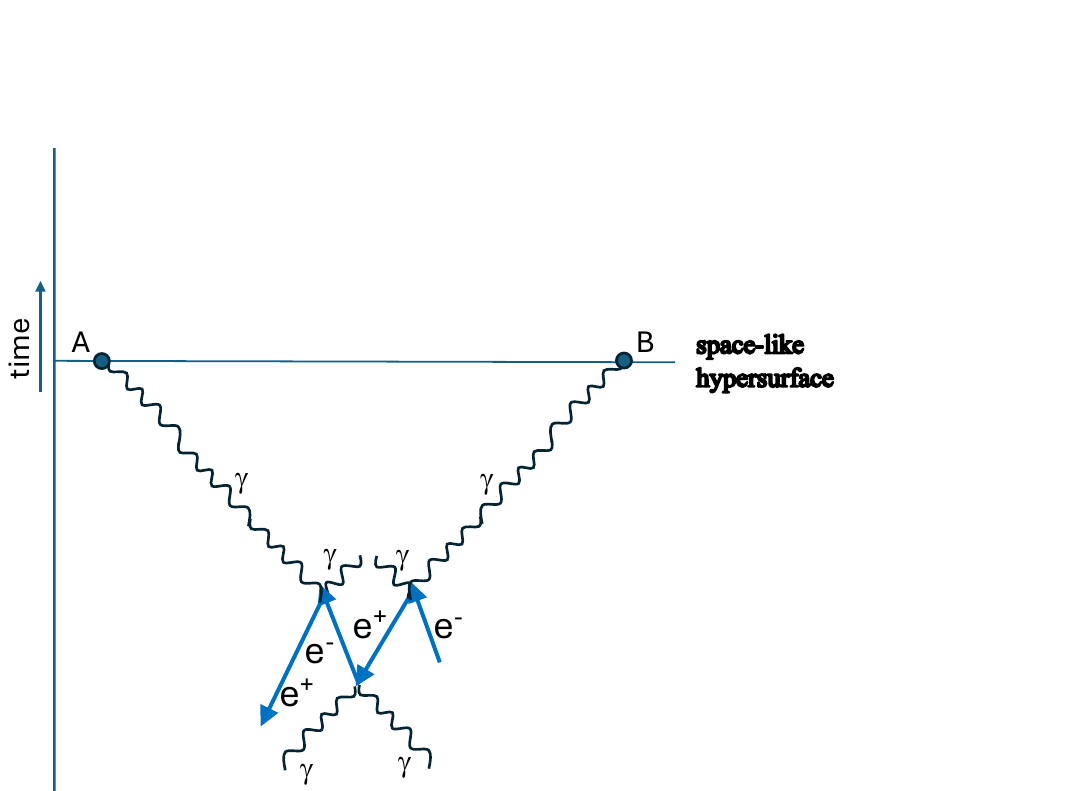}} \\
c) \resizebox{0.6\textwidth}{!}{\includegraphics[angle=0, trim = 0in  2in 4in 1in, clip]{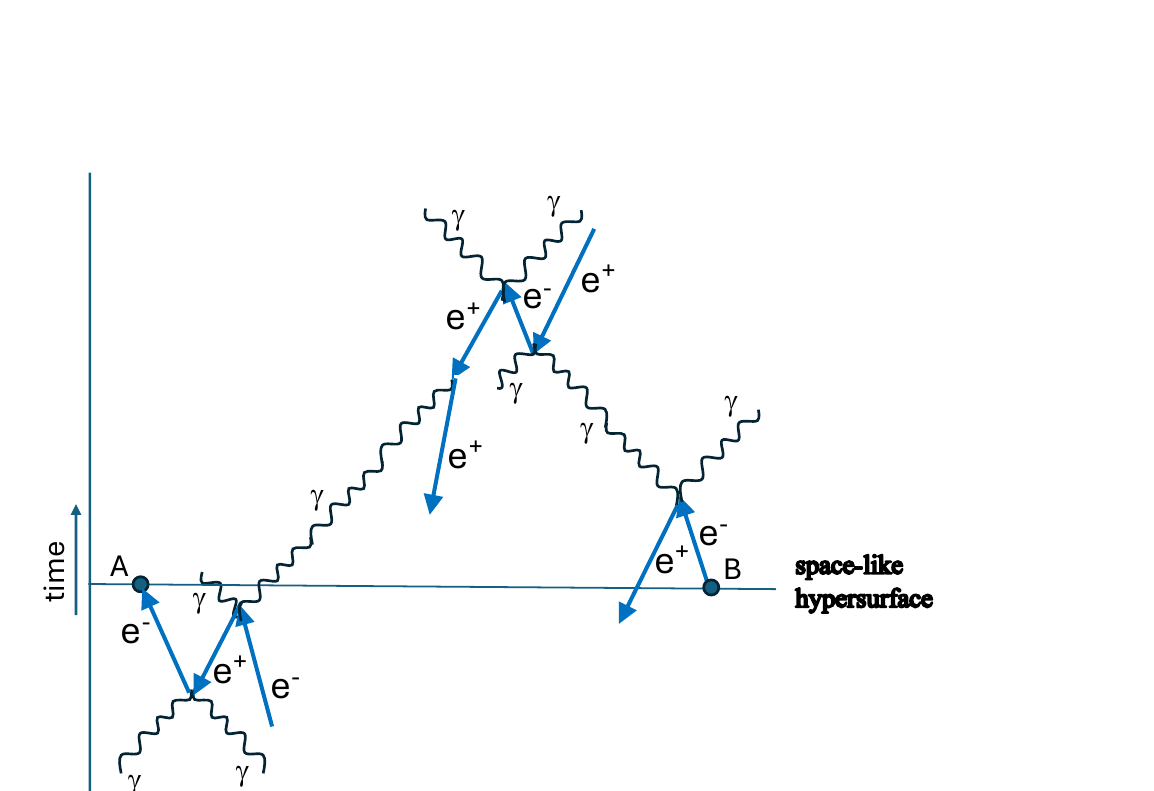}} 
\caption{Possible connections between partially entangled particles: a) successive pair production and annihilation partially entangles an electron and a positron with $n=5$ intervening interactions; b) two photons are partially entangled through a sequence involving $n=3$ intervening interactons; and c) two electrons are partially entangled through a sequence of $n=6$ interactions mediated mostly by exchange of photons. (Pair production and annihilation events shown with two photons include interactions of a single photon with a a static electromagnetic field.)}
\label{figpath}
\end{figure}

The probability of entanglement connection vs. distance is argued to be related to the distribution of free path lengths. A particle that has traveled in a vacuum with no interactions is more likely to be highly entangled with a particle near its point of origin than a particle that has had multiple interactions, each of which serves to spread any entanglement, and thus to dilute the entanglement of a given pair. Specifically, it is known that the entanglement of a pair of particles that is initially maximally entangled generally decays exponentially with the number of subsequent interactions (see Appendix \ref{AppendixA}). We take the number of interactions as a proxy, inversely, for the degree of entanglement.  Hence we choose a certain number of interactions $N_e$ and posit that two nodes are linked iff there is a sequence of particle paths connecting them involving $N_e$ or fewer interactions. We first  find the probability $P(l,n)$ that a particle emanating from point $\bx_A$ generates a series of particles, generally off the space-like hypersurface, through a sequence of $n$ interactions, the last particle arriving at point $\bx_B$ at a distance $l$ from $\bx_A$ , as in Fig. \ref{figpath}.  We assume that the entire process takes place in a uniform background, so that the normalised probability of the particle arriving at $\bx_B$ with no interactions is $P(l,0)= \lambda \exp(-\lambda l)$ where $\lambda$ is the mean free path. We integrate over the position 
$\bx_i$ of each interaction and find:
\begin{equation}
\label{eqPln}
P(l,n)= \int d\bx_1 \int d\bx_2 \ldots \int d\bx_n  s^{n-1} \exp [ -s (||\bx_1 - \bx_A|| +  ||\bx_2 - \bx_1||+\ldots+||\bx_B - \bx_n||)]
\end{equation}
a quantity which depends on $\bx_A$ and $\bx_B$ only through their distance $||\bx_B - \bx_A|| \equiv l$. The positions $\bx_i$ do not need to be on the space-like hypersurface. Some must be in the past, as in the case of an entangled  Bell pair emanating from a source. Others are in the future, as in delayed-choice entanglement swapping. The N-interaction distribution of path lengths is found to be \cite{refFeller}:
\begin{equation}
\label{Pdistln}
P(l,N)= 4\pi l^2 \left(\frac{3}{4\pi N \lambda^2}\right)^{3/2}\exp \left(-\frac{3l^2}{4N\lambda^2}\right)
\end{equation} 
For any given number of interactions $n$, $P(l,n)$ is a decreasing function of total distance $l$.
The probability of a connection between a particle initially at $\bx_A$ and one  at $\bx_B$ is $P_{AB}=\sum_{n=0}^{N_e} P(n | l)$. The summands $P(n | l)$ are the a posteriori probabilities that a particle path has  included exactly $n$ interactions, given that the total path traverses a distance $l$.  (We take it to be a consequence of Big Bang cosmology that there is {\it some} sequence connecting the particles for a sufficiently large number of past interactions.) The a posteriori probability is obtained from $P(l,n)$ using Bayes'  Rule:
\begin{equation}
\label{nBayes}
P(n | l) = \frac{P(l,n)P(n)}{\sum_{k=1}^\infty P(l,k)P(k)}
\end{equation}
Assuming no prior knowledge of the number of interactions, we take $P(k)$ to be flat, and find
\begin{equation}
\label{ngivenl}
P(n | l) \propto n^{-3/2} \exp \left(\frac{-3l^2}{4 n \lambda^2}\right)
\end{equation}
which is then normalized.
Summing over all $n \le N_e$ gives the desired cumulative  distribution:
\begin{equation}
\label{distNe}
P(n \le N_e | l) = \frac{\sum_{j=1}^{N_e} \left[j^{-3/2} \exp \left(\frac{-3 l^2}{4 \lambda^2 j}\right)\right] }{ \sum_{j=1}^{\infty} \left[j^{-3/2} \exp \left(\frac{-3 l^2}{4 \lambda^2 j} \right)\right] }
\end{equation}
which is the probability of a link from A to a node corresponding to a particle B at distance $l$ in our network model of partial entanglement. 
We next consider the behavior for realistic values of 
$\lambda$ and generalize to non-uniform distributions of matter.

\begin{figure}
\resizebox{1.0\textwidth}{!}{\includegraphics[angle=0]{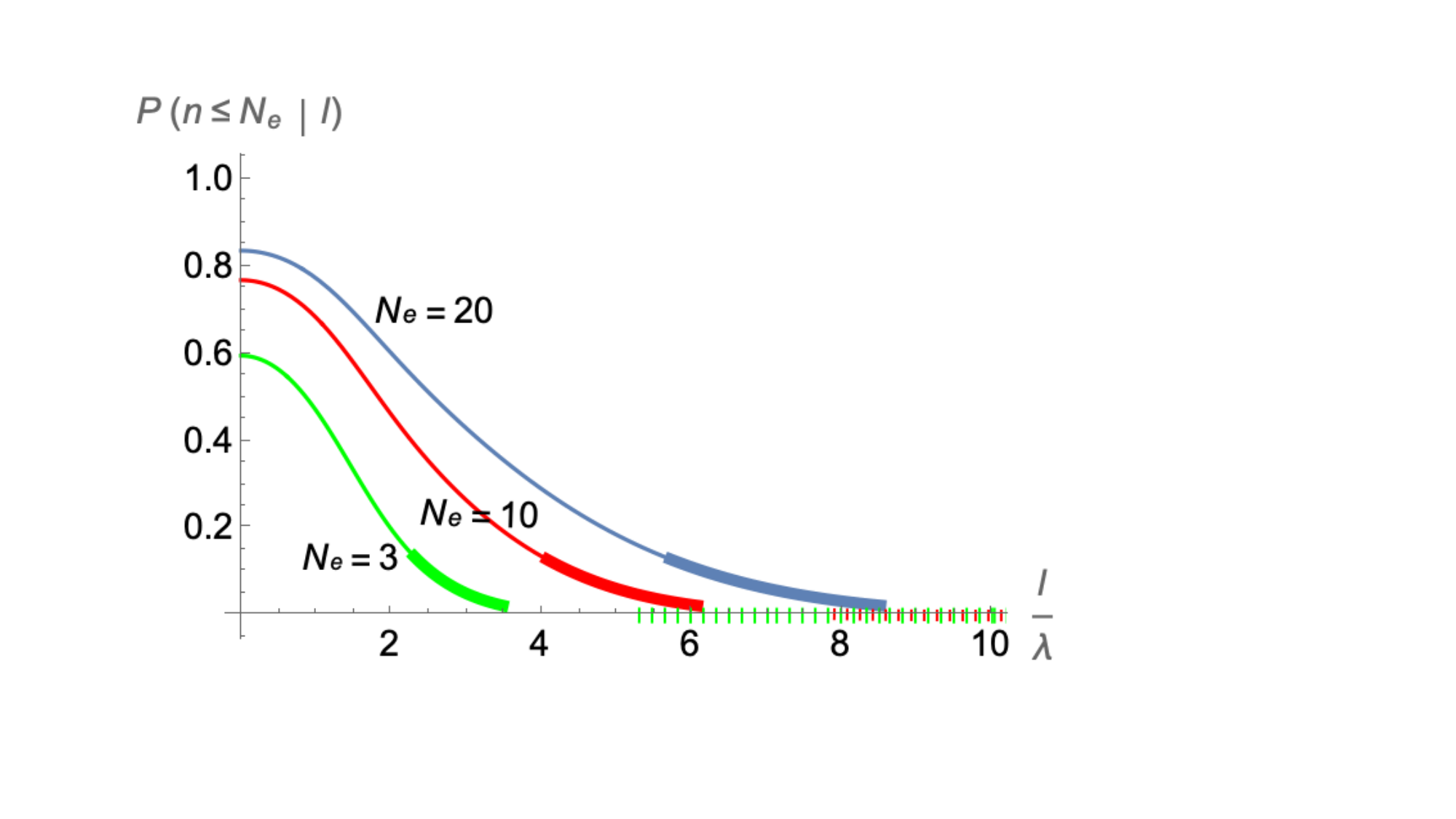}}
\caption{Probability of partial entanglement between two particles in a given medium, as defined by maximum number $N_e$ of allowed intervening interactions, vs. the distance $l$ between the particles in units of the mean free path $\lambda$, under the simplifying assumptions that all intervening particles have the same mean free path and that there is only one sequence connecting the particles. The entanglement connections define small-world networks if all distances are in the ranges marked by thick curves; they define random networks, in which clustering is lost, if the distances are smaller. For larger distances (dotted lines) small-world connectivity is lost.}
\label{figPvsl}
\end{figure}

\section{Entanglement Connection Pattern in the Universe vs. Small World Requirements}

The distribution (\ref{distNe}) is to be compared to Nguyen and Martel's requirement (\ref{Nguyenbound}) for small-worldedness, which also applies in our setting (see Appendix \ref{AppendixB}). But since
(\ref{Nguyenbound})  applies to power-law distributions, we need to generalise to any distribution that is bounded by  power-law distributions with exponents $r$ satisfying (\ref{Nguyenbound}).  In 3d space the requirement is satisfied if the connection probabilities fall off with distance $d$ faster than $1/d\,^3$ and slower than $1/d\,^6$.  
Our distribution (\ref{distNe}) inevitably fails the second test, for any choice of $\lambda$, since it falls exponentially as $l \rightarrow  \infty$, but by restricting to a finite range of $l$, there is still hope that the requirement may be satisfied in the restricted domain.  This is a natural approach in a cosmographic context, where particle mean free path varies tremendously depending on the region of space considered, and one can consider different regions separately, each of finite size.

Additionally, nodes defined as particles leave out all spatial positions that are not occupied by particles, so that physical distances in the network no longer correspond to lattice distances in the space of nodes as studied by Nguyen and Martel \cite{Nguyen}.  We begin by restricting attention to large scales in a finite region where the distribution of matter can be taken to be uniform. For constant density of particles at a scale of interest that is sufficiently large with respect to that density, lattice distance can indeed be approximated as physical distance.

Starting at the largest scale, let us imagine that the nodes are not individual particles, but galaxies, and that two galaxies are connected if they are joined by at least a single pair of particles that are entangled to a degree corresponding to $N_e$ intervening interactions or fewer. We add stray particles in inter-galactic space to the collection of nodes.  In this world, the mean free path  
can approach the size of the observable universe, depending on the type of particle (see Table \ref{tabMFP}). Using connections defined, say, by partial entanglement corresponding to $N_e=10$ interactions, the probability of entanglement (\ref{distNe}) falls off more slowly than $l^{-2k}=l^{-6}$ out to about $l/\lambda =6$, as seen in Fig. \ref{figPvsl}. However, for  $l/\lambda <4$, as holds for photon connections,  the probability falls off even more slowly than $l^{-2k}=l^{-3}$,  so in this regime  the network is not small-world, but random, without clustering.  The path lengths remain short, as in a small world.

The situation is similar at interstellar scales within a aingle galaxy.  High-energy photons have a mean free path of $\lambda = 10^{19}$m, which is comparable to the size of a typical galaxy.  So if we restrict attention to one galaxy at a time, connections composed of segments involving such photons again define either a small-world network or a random network among the nodes (particles) thus connected. An analysis in terms of the particles that are involved in typical interactions and their energies is needed to reach a rigorous conclusion.

Moving further downscale, a photon mean free path in interplanetary space within the solar system, $\lambda = 10^{13}$m is
somewhat larger than the diameter of the solar system, so the same argument applies.  For electrons, the mean free path $\lambda \approx 10^{10}$m is less, giving a probability of partial entanglement, say for $N_e=10$, that drops off with distance  more  slowly than $l^{-6}$, only within a small fraction of the solar system diameter. Small-world structure in this domain would again depend on photon connections. Again, a more precise accounting of types of interactions and the particles involved is needed.

The situation is different within a star, where free paths are of the order of centimeters or smaller, much less than the size of any domain naturally defined to include the entire star.  The small-world property can  only be maintained by initially choosing the distance $D$, within which all nodes are connected by definition, to include the star. The condition that the probability fall off faster than $r^{-3}$, to maintain clustering, is automatically satisfied for $r>D$, with D of typical stellar size.

\begin{table}[t]
\caption{Order-of-magnitude estimates of the mean free path of photons and electrons  in different regions of space, and the distances (maximum $l$) out to which partial entanglement mediated by photons or electrons give rise to small-world or random-network structure, for different definitions of partial entanglement  corresponding to different numbers $N_e$ of intervening interactions..} 
\label{tabMFP}
\begin{tabular}{c|cccc}
\hline
     & intergalactic space & interstellar space & solar system to& star interiors\\
      & in observable & in  galaxies &heliopause & (Sun type)\\
      & universe \cite{Peacock}&  \cite{Draine}& \cite{Kivelson}&\cite{Kippenhahn}\\
      &  $\le10^{27}$m & $ \le10^{19} $-$ 10^{21}$m & $\le10^{13}$m & $\le 10^{10}$m\\
  \hline
     &  \multicolumn{4}{c}{photons} \\
      \hline
mean free path & $10^{27}$m & $10^{19}$-$10^{22}$m & $10^{13}$m &  $10^{-4}$-$10^{-2}$m \\
 \hline
 max $l$ for small world &&&&\\
   $N_e=10$  & $10^{28}$m & $10^{20}$-$10^{23}$m & $10^{14}$m & $10^{-3}$-$10^{-1}$m   \\
   $N_e= 3$  & $10^{28}$m & $10^{20}$-$10^{23}$m & $10^{14}$m & $10^{-3}$-$10^{-1}$m   \\
   \hline
 max $l$ for random network &&&&\\
    $N_e=10$  & $10^{28}$m & $10^{20}$-$10^{23}$m & $10^{14}$m & $10^{-3}$-$10^{-1}$m   \\
   $N_e= 3$  & $10^{27}$m & $10^{19}$-$10^{22}$m & $10^{13}$m & $10^{-4}$-$10^{-2}$m   \\
  \hline
 \hline 

  \hline
     &  \multicolumn{4}{c}{electrons} \\
  \hline
mean free path & $10^{19}$m & $10^{12}$-$10^{15}$m & $10^{10}$m &  $10^{-10}$-$10^{-8}$m \\
 \hline
 max $l$ for small world &&&&\\
   $N_e=10$  & $10^{20}$m & $10^{13}$-$10^{16}$m & $10^{11}$m & $10^{-9}$-$10^{-7}$m   \\
   $N_e= 3$  & $10^{20}$m & $10^{13}$-$10^{16}$m & $10^{11}$m & $10^{-9}$-$10^{-7}$m   \\
   \hline
 max $l$ for random network &&&&\\
   $N_e=10$  & $10^{20}$m & $10^{13}$-$10^{16}$m & $10^{11}$m & $10^{-9}$-$10^{-7}$m   \\
   $N_e= 3$  & $10^{19}$m & $10^{12}$-$10^{15}$m & $10^{10}$m & $10^{-10}$-$10^{-8}$m   \\
  \hline
  
\end{tabular}
\end{table}

Thus we have a preliminary argument for small-world structure or random network structure in the universe if we restrict attention to scales larger than the sizes of typical stars, and conjoin the different network structures for regions with different matter densities. Such a conjunction, which in varying the size of the ``universe" only varies the size of the intergalactic region containing varying number of similar galaxies, is easily seen to give the required logarithmic dependence of average path length on ``universe" size. That is because path lengths in the intergalactic region, considered separately, already have this property, and path properties within galaxies are not affected by varying their number.  

\begin{figure}
a) \resizebox{0.8\textwidth}{!}{\includegraphics{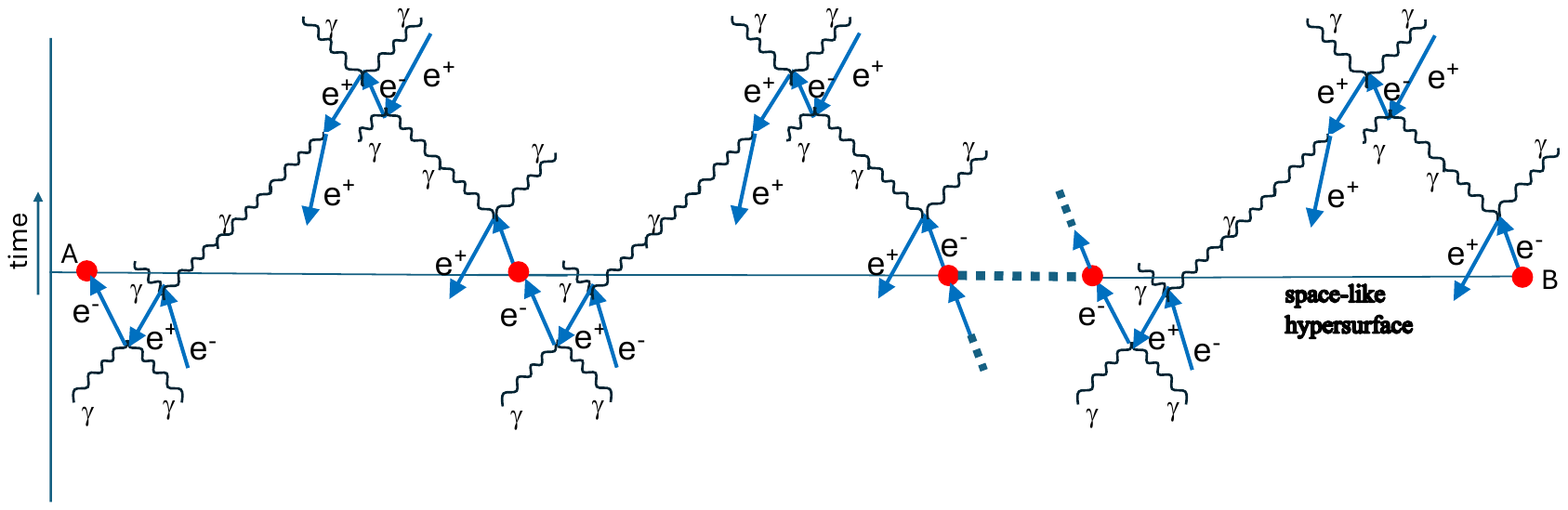}} \\
b) \resizebox{0.8\textwidth}{!}{\includegraphics{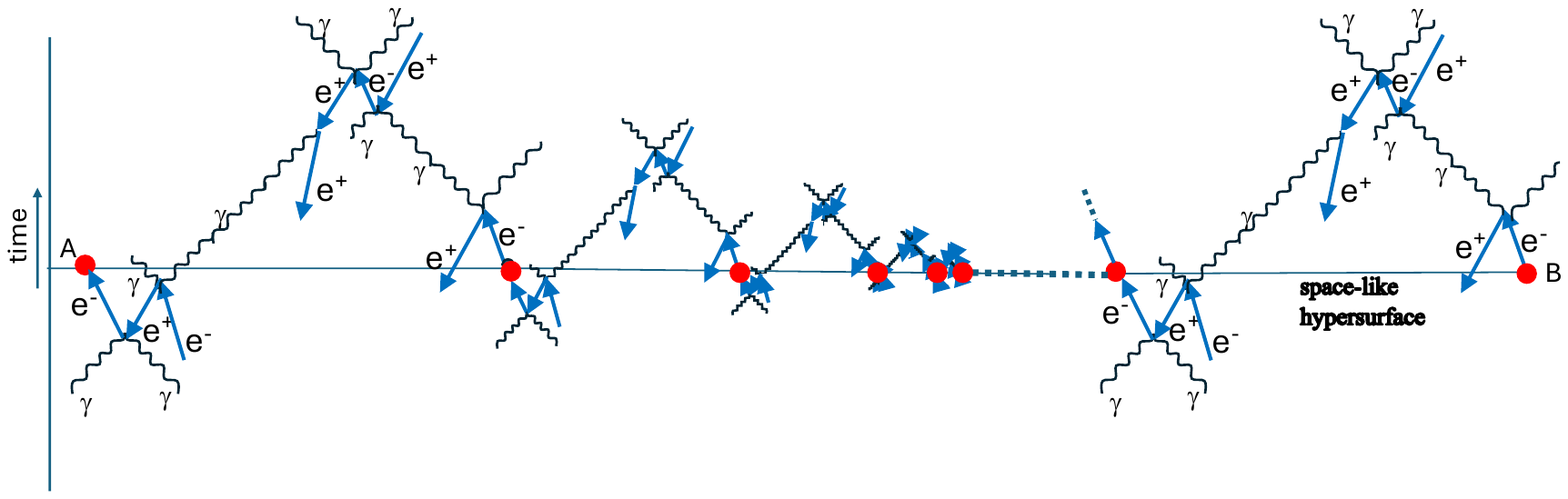}}
\caption{a) Chain of links between particles (red dots), on a given space-like hyper-surface, connecting two particles A and B, each link defined by a sequence of less than $N_e$ interactions giving partial entanglement at each link.  (For convenience, the same sequence of interactions is shown to mediate each link.) b) A similar chain of links in a fractal medium in which mean free paths shorten as matter density increases intermittently, giving proportionately shorter trajectory segments and shorter links.  If the descent in scale continues indefinitely, an arbitrarily long chain of links between A and B is required, and the small world property is lost.}
\label{figlinkchains}
\end{figure}

A more systematic view of the matter distribution in the universe is that the distribution is fractal \cite{refPietronero}. In general fractal structure  and small world structure have been found to be complementary patterns  \cite{refCsanyi}, that only co-exist on a limited range of scales, in agreement with our analysis. The proof of small world structure in Appendix \ref{AppendixB} can be seen to break down for fractal matter distributions. Indeed if one were to extrapolate the above analysis of cosmography to smaller and smaller scales, for reasonably smooth scaling of mean free path, the number of links required to connect two nodes would generally increase without limit, as suggested in Fig. \ref{figlinkchains}b. It is interesting that links defined by partial entanglement of a given degree only contribute to small world structure on scales describing largely empty space, where links of a more familiar nature found within stars and planets are absent.

\section{Implications}

A more detailed analysis is needed both in regard to the spectrum of interacting particles and in regard to cosmography.  The analysis in Section \ref{sec2} must be generalised to multi-particle trajectories, with each type of particle characterised  by a set of mean free paths, one for each type of interaction that can terminate the path (including spontaneous decay), also taking account of energy-dependence.  All types of particles need be considered. Heavy leptons are unstable, with short total mean free paths, and play a negligible role in long-range connections. The same is true for the unstable bosons of the Standard Model. At the other extreme, neutrinos in intergalactic space have mean free paths much larger than the observable universe and do not contribute. But quarks confined in protons or in light atoms  have total mean free paths comparable to the size of the observable universe.  Those need be considered in an analysis of multi-particle trajectories that would apply the methods of this preliminary study  in an extended treatment.  In regard to cosmography, the notion of uniformly empty intergalactic space must be replaced by the network of filaments and sheets that constitutes the large-scale cosmos.

All considered, however, the conclusion that entanglement gives small-world or random network structure in the observable universe on large scales is expected to persist.  Protons and light atoms  play a role similar to that of photons and electrons in our study, while other particles will contribute very short trajectory segments that do not affect the overall picture. The density of matter in filaments and sheets is less than that in typical galaxies, by many orders of magnitude.  Thus our conclusions about mean free paths and entanglement network structure in intergalactic space are not affected. We expect to find again that entanglement in an extended treatment must be supplemented by other types of connections on stellar or planetary scales - crudely by an appropriate choice of the parameter $D$ or using some more detailed definition of small-scale connections. 

One could also seek small-world or random network structure in the entire universe, to include the large, possibly infinite, unobservable portion.  Because of cosmological expansion, any entanglement with particles deep in the unobservable universe must be rooted in the distant past.  So the full 4-dimensional geometry must be considered. The impact of entanglement with unobservable particles and of the resulting network structure on  observable phenomena is obscure.

Indeed, the impact of entanglement-mediated small-world structure within the observable universe also remains to be elucidated.
In the classical realm, small-world structure is known to facilitate, but not to guarantee synchronisation of dynamical systems placed at the nodes \cite{BarahonaPecora}. Entanglement itself is not readily cast as synchronisation, except in some non-standard deterministic interpretations of quantum mechanics \cite{DuaneEntropy}.  But entanglement can propagate synchronisation and enlarge the synchronised domain. Two systems can synchronise if  each is classically connected to one of a pair of entangled particles, thus extending the effects of entanglement, a phenomenon that has been explored primarily in a computer science
context \cite{refChuang,refQuan}. Certainly stars and planets are rife with synchronised phenomena \cite{Sync}.  If one envisions entanglement on a cosmic scale extending  synchronisation beyond local domains, then 
there is a possibility that the entire observable universe, or even the larger universe, is constrained in a manner not heretofore appreciated.  Toward such understanding, further development of the picture of a small world based on a combination of large-scale entanglement and smaller scale classical connections is indicated.

\appendix

\section{Degree of Partial Entanglement From Negativity}
\label{AppendixA}

Partial, rather than full entanglement is used to define network connections because a particle can be fully entangled with only one other particle, precluding a chain of such connections. The degree of entanglement of two particles $A$ and $B$  can be defined in terms of the {\it negativity} of the density matrix $\rho$ describing a statistical ensemble of quantum states encompassing the two particles. Specifically, one considers the {\it partial transpose} $\rho^{T_B}$ with respect to one of the two subsystems, say $B$, and defines the negativity $\mathbb{N}$ as the sum of the magnitudes of the negative eigenvalues of $\rho^{T_B}$,  i.e.
\begin{equation}
\label{defnegativity}
\mathbb{N}  \equiv \sum_{\lambda_i < 0} |\lambda_i|
\end{equation}
where $\lambda_i$ is the ith eigenvalue of $\rho^{T_B}$.
The partial transpose denoted by the superscript $T_B$ is the transpose only with respect to the indices belonging to particle $B$.  That is, for a matrix $M$ that is indexed by the quantum numbers of the composite $A,B$ system, the partial transpose $M^{T_B}$  has elements given by $M^{T_B}_{i^A i^B, j^A j^B} = M_{i^A j^B, j^A i^B}$  where $i^A$ and $j^A$ each denote the collection of quantum number indices that pertain only to $A$, while  $i^B$ and $j^B$ denote the collection of quantum number indices that pertain only to $B$.  ($M$ can be envisioned as a matrix of blocks indexed by $i^B$ and $j^B$ with each block indexed internally by  $i^A$ and $j^A$.) 

The relationship of negativity to entanglement stems from the observation of Peres \cite{refPeres} that positivity of the eigenvalues of 
the partially transposed joint density matrix $\rho_{AB}^{T_B}$ is guaranteed if  $A$ and $B$ are separable. The partial transpose with respect to $B$ gives time reversal for the $B$ subsystem but not the $A$ subsystem \cite{refPeres}.   The resulting state is physically realisable - with a calculable density matrix - if $A$ and $B$ are separable but is physically impossible if they are entangled.  In the latter case the difficulty is reflected in the negative probabilities that are the eigenvalues of $\rho^{T_B}$ and can be quantified as the sum of those negative eigenvalues \cite{refVidalWerner}.

In general, negativity depends exponentially on the number of intervening interactions $N$:
\begin{equation}
\label{entangledecay}
\mathbb{N} = \exp(-N/\nu)
\end{equation}
where the constant $\nu$ is a number of interactions that characterises the entanglement decay rate. The behaviour (\ref{entangledecay}) is in agreement with the intuition that the number of intervening interactions gives the degree to which entanglement is spread to other particles. It is assumed that connections are dominated by particles in their ground states, so that ``entanglement sudden death"  resulting from spontaneous emission due to vacuum interactions \cite{YuEberly} need not be considered. An exponential relationship similar to (\ref{entangledecay}) was first put forward for negativity as a measure of entanglement in a many-body context \cite{refAudenaert}. The relationship was also equivalent to  previous results for the decay of {\it concurrence} (absolute value of the single negative eigenvalue of $\rho^{T_B}$ ) for particles in a Bell pair undergoing a series of interactions before measurement \cite{refZyczkowski}.  We extrapolate from these results to use $N$ as a proxy for negativity.

\section{Conditions for Small-World and Random Network Structure Where Connection Probability Decreases 
With Distance}
\label{AppendixB}

In a network with a distance metric $d(A,B)$ that is prescribed {\it a priori}, he network structure is {\it random} if the probability of connection between any two nodes is independent of distance.  Small-world structure results if the dependence on distance has properties given by Nguyen and Martel \cite{Nguyen}, for the case where there is exactly one ``outgoing" connection, or any fixed number of such connections, at each node.  Here we sketch Nguyen and Martel's proof and then sketch a generalisation  to the case where any number of long-range connections is allowed at each node.

We consider connection structures in which nodes in some specified neighbourhood of one another are always connected and the probability $P(A,B)$ that there is a long-range link between more distant nodes $A$ and $B$ is given by:
\begin{equation}
\label{Pdist}
P(A,B) = \frac \kappa {[d(A,B)]^{-r}}
\end{equation}
where $r$ is a decay rate parameter of interest and $\kappa$ is a constant.  If the number of long-range links emanating from a given node is fixed, as e.g. in \cite{Nguyen}, then $\kappa$ is given by a normalisation condition.  Otherwise $\kappa$ is a free parameter.  The average number of outgoing links from a given node $u$ is given by:
\begin{equation}
\label{avgq}
<q_u>=\kappa \sum_{v \ne u} \frac 1 {(d(u,v))^r}
\end{equation}
In the following, we consider various ranges of values of the decay parameter $r$.

\subsection*{Small-World Diameters for Networks With $r<2k$}  

{\it Fixed number $q$ of outgoing links from each node:} The probability (\ref{Pdist}) is re-defined as the probability that a given one of the  $q$ links emanating from $A$ ends at $B$.  An undirected link can be interpreted as a directional link by arbitrarily assigning one of the end nodes as the ``owner".  Following \cite{Nguyen}, we focus on the 1-dimensional case $k=1$.  We consider successive decomposition in a branching of the entire 1D space of
$n$ nodes into $n^{1-\xi}$ segments of length $n^\xi$,  for some $\xi$ with $0<\xi<1$, a decomposition of each of those segments into smaller segments each of length
$n^{2\xi}$, and so on, until we reach a final decomposition into segments of some minimum length $x_o$, at the smallest branches of the tree. It was shown, under the assumptions of \cite{Nguyen}, that if  $x_o$ is not too small, then the decomposition at each stage has the following Property A: There is a single link joining any two segments
(i.e.  linking some pair of members of the respective segments) with probability that approaches unity as $n$ becomes large. This results in a doubling of the number of links needed to connect any two nodes in a branch, i.e. its diameter,  as one moves down the tree toward bigger branches.  One might therefore imagine that the diameter of the tree trunk would depend  logarithmically on the size $n$. But the sequence of segment sizes $(x_o,x_o^{1/\xi},x_o^{1/\xi^2},x_o^{1/\xi^3},\ldots)$ is double exponential - not geometric - resulting in a network diameter that varies as a polynomial of the logarithm of network size 
\begin{equation}
D(n)=c(\log n)^\beta
\label{Dnetwork}
\end{equation}
where $\beta$ depends only on $r$ and $\xi$, with a proportionality constant $c$ that depends on the overall connectivity parameter $\kappa$ as well, as shown in detail in \cite{Nguyen}.

The key step to establish the poly-log dependence of network diameter on network size $n$ is thus to derive property A, which is then applied at each step in the recursion. If there are exactly $q$ links emanating from each node, as in \cite{Nguyen}, the probability that a link from a given node in a segment $S$ of length $x^\xi$ goes to a given node in another segment $T$ of the same size (same level in the branching) is $\ge \kappa/x^r$, since the nodes are separated by a distance at most $x$. (We use lattice distance as a measure of physical distance in a uniform medium - see text.) Noting that $|T|=|S|= x^\xi$, we have that the probability that the link goes to {\it any} node in $T$ is $\ge \kappa x^{\xi-r}$

The probability that the link goes to {\it no} node in $|T|$ is $\le 1 - \kappa x^{\xi-r}$, while the probability that {\it none} of the $q$ links from the given node go to any node in $T$ is $\le (1- \kappa x^{\xi-r})^q\le \exp(-q\kappa x^{\xi-r})$, the last step following from the algebraic identity $1-A \le \exp(-A)$. Considering now {\it all} the nodes in $S$, each with $q$ long-range links, the probability that {\it none} of the links from $S$ go to any node in $T$ is $\le \exp(-q\kappa x^{(\xi-r)}|S|)=\exp(-q\kappa x^{2\xi-r})$. The probability of the complementary event, that at least one link goes from $S$ to $T$ is therefore
\begin{equation}
\begin{aligned}
 &P(\text{there is at least one link from some node in } S \text{ to some node in } T) \\
 &\quad \ge 1 - \exp(-q\kappa x^{2\xi-r})
\end{aligned}
\label{PStoT}
\end{equation}

For a deep enough tree with the smallest branches no smaller then $x_0$, most of the recursion steps will involve large values of $x$ and $P \approx 1$, provided $r<2$,  since $\xi < 1$, as required for construction of the tree, and we can choose $\xi > r/2$.  That was the basis of the proof in \cite{Nguyen} that almost all branch points exhibit Property A and that the diameter of the entire network of $n$ nodes has the poly-log form $\propto \log^\beta n$. 
 
 {\it Fixed probability of outgoing links from a given node to distant nodes without restriction on their number:}  We generalise the previous analysis of networks with exactly one long-range link at each node by Nguyen and Martel \cite{Nguyen}  to networks with arbitrary $\kappa$ and the number of links at each node given probabilistically by (\ref{Pdist}).  The change does not affect the reasoning greatly. The probability that a given node in $S$ is connected to a given node in $T$ is $\ge \kappa/x^r$. The probability that it is {\it dis}connected is $\le 1 - \kappa/x^r \le \exp(-\kappa x^{-r})$. The probability that all nodes in $|S|$ are disconnected from all nodes in $|T|$, with both segments of size $x^\xi$ is $\le \exp(-\kappa x^{2\xi-r})$, and the probability of the complementary event
 - a connection between $S$ and $T$ -  is $\ge 1 - \exp(-\kappa x^{2\xi-r})$. The rest of the argument is as before.

 The value of $\kappa$, can be absorbed in a re-defined ``normalisation constant", since neither the value nor the interpretation of that constant was used in the proof in \cite{Nguyen}\footnote {The parameter in \cite{Nguyen} that we will adjust is $C_r$ where $1/(2C_r)$ is a lower bound on $\kappa$.  The specific  definition $C_r \equiv \sum_{i=1}^\infty i^{-r}$ was not used except to establish that bounding relation.}. $\kappa$ enters only in setting the minimum segment size $x_0$ at the smallest branches.  The diameter of these smallest segments -which also must be large enough so that the uniformity assumption applies - enters only as a multiplicative factor in the sequence of diameter doublings as we descend the tree, and thus the re-defined normalisation constant $\kappa$ only affects the poly-log form (\ref{Dnetwork}) in the proportionality constant $c$.  The poly-logarithmic dependence on network size that implies a small world persists.
 
 We must consider the possibility that a node $u$ is completely disconnected, with $q_u=0$, as occurs with probability $\prod_{v \ne u}[1-  \kappa /d(u,v)]$. In that case the diameter of any segment containing $u$ is formally infinite.  If we take the strong position that partial entanglement, appropriately defined, gives at least one long-distance outgoing link from any particle, then the proportion of disconnected nodes must be negligible. If we take the weaker position that there is a small but finite proportion of disconnected nodes, we can rely on the usual modification of the definition of small-world networks to include situations where there is a ``Giant Connected Component", itself a small world, that remains when disconnected nodes are ignored \cite{refGCC}.

    Both Nguyen and Martel's original construction and our extension generalise from the 1D case described above to k-dimensional space by replacing the segments with hypercubes. The requirement that $r<2$ becomes $r<2k$  \cite{Nguyen, NguyenExt}.
    
\subsection* {Random Network Structure With Loss of Clustering for $r<k$}
In k-dimensional space, it is easily seen that for a network with $r<k$, the clustering property that is part of the definition of small-world network structure is lost (as it is for the $q=1$ case considered in \cite{Nguyen}). In such a network, the average length of a link from some node $u$ to any other node in a space where nodes are distributed homogeneously, can be approximated as:
\begin{eqnarray}
\label{avglink}
\sum_v d(u,v) P(u,v) &\approx& \int_V d^k x \frac \kappa {||x||^r} \\    \nonumber
                                 &\propto& \int_0^{R_{max}} dR \;R^{k-1}  \frac 1 {R^r}    \\   \nonumber
                                 &=& \int_0^{R_{max}}  dR\; R^{k-r-1}                          
 \end{eqnarray}
The integral is seen to diverge as the volume $V$, or its maximum radius $R_{max}$ increases, if $r<k$. Nodes are clearly not clustered. Formally, two nodes that are linked to $u$ are likely to be separated by a distance of the order of the network size, with a low probability that they themselves are linked, thus failing the requirement for clustering that two nodes linked to a common node are likely linked to each other. Thus even though the diameter of a network with $r<k$ is as small or smaller than the diameters of the networks with $k<r<2k$,  networks in the former category are random and not small-world.
 
\subsection*{Large Worlds for $r>2k$}
For a network with $r>2k$, we generalise the argument in \cite{Nguyen} for $k=1$ to arbitrary $k$, to demonstrate large-world structure, characterised by a diameter that varies as a power of network size, rather than logarithmically.  For $r>2k$, we can choose $\gamma$ such that $1/(r/k\;\;-1) < \gamma <1$. The probability that the length of any one link is less than $n^{\gamma/k}$ is
\begin{eqnarray}
\label{Pshortlink}
P(d(u,v)<n^{\gamma/k}) &=& 1 -\kappa \sum_{v:d(u,v)>n^{\gamma/k}} d(u,v)^{-r} \\             \nonumber
                                 &\approx& 1 - O( \int_{n^{\gamma/k}}^{n^{1/k}} dR\; R^{k-1}  \frac \kappa {R^r})   \\.     \nonumber
                                 &=& 1- O(n^{-\gamma (r/k-1)})
\end{eqnarray} 
since the lower integration bound dominates for large $n$.                                
The probability that all $<q>n$ links are less than $n^{\gamma/k}$ must then be 
$\ge 1 - <q> n O(n^{\gamma (1- r/k)}) = 1 - <q> O(n^{1-\gamma (r/k - 1)}) $.  Since $\gamma > 1/(r/k-1)$, the exponent of $n$ is negative and the probability is bounded below by a number that approaches unity as $n$ becomes large.  Thus almost all links
are of length less than $n^{\gamma/k}$.  The diameter of the network must therefore be at least $n^{1/k}/n^{\gamma/k} = n^{(1-\gamma)/k}$, a positive power of the network size, as was to be shown.

\end{document}